\documentclass[trackchanges]{aastex701}

\graphicspath{{figures/}}
\begin{document}

\title{Forward Modeling of Dust-Induced Stray Light in Ground-Based Coronagraphs: A Dual-Path Monitoring Approach for High-Precision Inner Corona Observations}

\author{Xiande Liu}
\affiliation{Yunnan Observatories, Chinese Academy of Sciences, Kunming 650216, China}
\affiliation{School of Physics and Astronomy, Yunnan University, Kunming 650091, China}
\email{liuxiande@ynao.ac.cn}

\author{Xuefei Zhang}
\affiliation{Yunnan Observatories, Chinese Academy of Sciences, Kunming 650216, China}
\email{zhangxuefei@ynao.ac.cn}

\correspondingauthor{XueFei Zhang}
\email{zhangxuefei@ynao.ac.cn}

\author{Yu Liu}
\affiliation{School of Physical Science and Technology, Southwest Jiaotong University, Chengdu 611756, China}
\email{lyu@swjtu.edu.cn}

\author{Tengfei Song}
\affiliation{Yunnan Observatories, Chinese Academy of Sciences, Kunming 650216, China}
\email{stf@ynao.ac.cn}

\author{Mingyu Zhao}
\affiliation{Yunnan Observatories, Chinese Academy of Sciences, Kunming 650216, China}
\email{myzhao@ynao.ac.cn}

\author{Mingzhe Sun}
\affiliation{Shandong Key Laboratory of Space Environment and Exploration Technology, Institute of Space Sciences, 
School of Space Science and Technology, Shandong University, Weihai 264209, Shandong, China}
\email{sunmingzhe2003@126.com}

\author{Feiyang Sha}
\affiliation{School of Physical Science and Technology, Southwest Jiaotong University, Chengdu 611756, China}
\email{shafy@my.swjtu.edu.cn}

\author{Jun Fang}
\affiliation{School of Physics and Astronomy, Yunnan University, Kunming 650091, China}
\email{fangjun@ynu.edu.cn}

\correspondingauthor{Jun Fang}
\email{fangjun@ynu.edu.cn}

\begin{abstract}
High-precision ground-based observations of the inner corona ($1.05$--$2.0 \, R_\odot$) are fundamentally constrained by instrumental stray light, particularly the additive background from dynamic dust accumulation on the objective lens. To address this issue,  we propose a correction method for the Spectral Imaging Coronagraph (SICG) based on \textit{dual-path real-time monitoring} and \textit{forward physical modeling}. By simultaneously imaging the objective lens surface, we obtain deterministic prior information on dust distribution. We construct a physical Point Spread Function using optical defocus parameters and reconstruct the non-uniform scattering background via convolution. Model parameters are retrieved through data-driven inversion constrained by polar coronal holes.

The method demonstrates excellent robustness under varying contamination conditions. After correction, the Root Mean Square noise in the polar background is reduced by approximately 67\% on average, and the Signal-to-Background Ratio improves by a factor of up to 3.7 under heavy contamination conditions. Comparisons with space-based SDO/AIA observations indicate that the corrected images recover the morphological structures of streamers with high fidelity. Further radial intensity analysis reveals that the correction process successfully restores the \textit{hydrostatic exponential decay} characteristic of inner coronal radiation. The fitted decay coefficient corresponds to a plasma temperature of approximately 2.0 MK, consistent with the characteristic formation temperature of the \ion{Fe}{14} 530.3 nm line. These results demonstrate that the method effectively eliminates the dominant systematic bias in ground-based observations, providing a reliable data foundation for high-precision coronal thermodynamic and dynamic research with the SICG.
\end{abstract}


\section{Introduction} 
\label{sec:obs}

The solar corona, representing the outermost layer of the solar atmosphere, consists of tenuous plasma with temperatures exceeding one million Kelvin. Its magnetic topology, electron density distribution, and associated dynamic processes---such as Coronal Mass Ejections (CMEs), coronal wave propagation, and the formation and evolution of streamers---play a decisive role in driving solar eruptive activities and the generation of space weather \citep{Parker1958, Aschwanden2005}. Consequently, high-precision observations of physical processes within the inner corona are critical for understanding the initiation of CMEs, the release of magnetic energy, and the origins of the solar wind.

However, in the visible spectral range, the brightness of coronal emission is only approximately $10^{-6}$--$10^{-9}$ times that of the solar photosphere \citep{vandeHulst1950}. This extreme brightness contrast renders high signal-to-noise ratio (S/N) observations technically challenging, making data quality highly dependent on the effective occultation of intense photospheric radiation and the rigorous suppression of instrumental stray light. To address this challenge, since the invention of the internally occulted coronagraph by \citet{Lyot1939}, ground-based coronagraphs have established themselves as essential tools for studying the inner corona. They continue to provide crucial observational constraints for research on CMEs initiation, stable streamer structures, and coronal wave propagation.

Although modern space-based coronagraphs (e.g., SOHO/LASCO and STEREO/COR) are capable of continuous monitoring over a wide field of view, and recently developed balloon-borne missions---such as the BITSE mission \citep{Gopalswamy2021} and the 50 mm white-light coronagraph experiment \citep{Kang2025}---have demonstrated the superiority of acquiring high-purity coronal signals in the near-space environment, these platforms are often constrained by telescope aperture and spectral configuration, resulting in limitations for the detailed diagnosis of the inner corona. In contrast, ground-based internally occulted coronagraphs, leveraging their large light-gathering apertures, high spectral resolution, and flexible observing modes, maintain unique and irreplaceable advantages for physical research in the inner coronal region ($r \lesssim 1.5 \, R_\odot$) \citep{Lin2004, Tomczyk2008}.

In particular, significant progress has been achieved in recent years in data analysis based on ground-based spectral imaging observations. Relevant studies have not only revealed the complex evolutionary characteristics of coronal emission across temporal and latitudinal dimensions \citep{Oloketuyi2025} but also realized breakthrough monitoring of the long-term evolution of the Sun's global coronal magnetic field \citep{Yang2024}. These cutting-edge results further underscore the urgent need for acquiring high-quality, low-stray-light ground-based coronal observational data to deeply investigate key problems in solar physics.

Despite meticulous optical design, ground-based coronal observations remain fundamentally constrained by stray light. In addition to the external sky brightness, instrumental stray light constitutes a critical bottleneck limiting data quality. Classical internally occulted coronagraphs effectively suppress diffracted light from aperture edges and ghost images from internal reflections through the implementation of Lyot stops and Lyot spots \citep{Newkirk1963}. 
 Nevertheless, as the primary optical element directly exposed to the full solar flux, the objective lens makes its surface cleanliness the final frontier in stray light control. Unlike the static scattering caused by the micro-roughness of optical surfaces, dust accumulation on the objective lens is inherently dynamic, varying in real-time with atmospheric conditions, wind direction, and maintenance schedules. These micron-sized dust particles act as strong scattering centers, superimposing a complex, time-varying, and non-uniform background across the field of view \citep{Fineschi1993, Elmore2000}.
In particular, for modern ground-based coronagraphs aiming for high-precision magnetic field measurements (e.g., the COSMO K-Coronagraph), \citet{deWijn2012} identified dust scattering on the objective lens as the primary environmental factor limiting instrument sensitivity and polarimetric accuracy. This additive stray light not only significantly degrades image contrast, obscuring faint coronal fine structures such as nascent CMEs or streamer boundaries, but also introduces severe systematic errors in photometric calibration, thereby directly compromising the accurate inversion of key physical parameters like electron density and temperature.

Traditional calibration methods exhibit significant limitations in addressing the issue of dust scattering from the objective lens. Standard flat-field correction is primarily intended to correct for pixel response non-uniformity and optical vignetting, functioning essentially as a multiplicative correction. However, stray light caused by dust scattering is superimposed onto the coronal signal as an additive component, which cannot be eliminated through flat-field division \citep{Kuhn1991}. While frequent physical cleaning of the objective lens can remove dust, this approach poses high operational risks in practical observations---particularly at remote or automated stations---and is incapable of addressing the dynamic accumulation of dust during data acquisition.

To extract pure K-corona signals from coronal observations, various background separation techniques have been developed. For instance, \citet{Shestov2014}, in processing solar eclipse data, utilized radial basis function fitting to separate the background based on the assumption that the spatial distribution of the F-corona is relatively smooth. However, this ``blind'' separation method, predicated on morphological assumptions, is challenging to apply directly to the correction of dust on the objective lens of ground-based coronagraphs. This is because the objective dust is in a defocused state, and the resulting scattering spots (Point Spread Function) on the focal plane, while diffuse, often retain specific geometric shapes and intensity gradients. Blind smoothing filtering is prone to confusing dust artifacts with the fine structures of coronal streamers, leading to the distortion of scientific signals.

Addressing the specific dust contamination issue inherent to ground-based instruments, existing research has primarily proceeded along two technical avenues. One approach adopts an empirical differential imaging strategy, which estimates and isolates the scattering background component by subtracting a ``relatively clean'' reference image from the contaminated scientific observation \citep{Sha2023_Acta}. In practice, such reference frames are typically acquired immediately after manual cleaning of the objective lens, with varying degrees of cleanliness achieved through repeated cycles of ``contamination--cleaning--re-observation.'' Although this method is conceptually intuitive and simple to implement, its effectiveness is highly contingent upon the consistency of observing conditions, instrument status, and scattering characteristics between the reference and science images. For ground-based coronal survey observations requiring long-term, continuous operation, frequently interrupting observations to obtain clean reference frames is often impractical.

Another category of methods attempts to construct empirical models, parameterizing the scattering background as a function of the integrated dust intensity extracted from auxiliary images of the objective lens \citep{Sha2023_SolPhys}. By performing differential fitting on multiple datasets with varying contamination levels, this method establishes a quantitative relationship between the scattering background and dust intensity, offering a degree of flexibility in engineering applications. However, such models rely essentially on statistical fitting; their parameters are not directly derived from a physical description of the optical scattering process but are instead coupled to the data characteristics of a specific instrument during a specific observing period. This empirical dependence not only limits the transferability of the model across different instruments or observing conditions but also affects its reliability when applied outside the calibration conditions.

To fundamentally overcome these limitations, we propose a stray light correction method based on \textit{dual-path real-time monitoring and forward physical modeling}. Leveraging the unique dual-path optical design of the SICG, we can image the objective lens in real-time while simultaneously acquiring coronal spectral data, thereby obtaining precise information on the positions and relative scattering cross-sections of dust particles. This key hardware innovation provides decisive prior information for the calibration algorithm, resolving the inherent degeneracy associated with the uncertainty of the Point Spread Function (PSF) in blind deconvolution techniques.

Under this framework, instead of relying on smoothness assumptions or empirical statistical relationships, we directly utilize the real-time monitored dust distribution as the scattering source function. By combining this with the instrument's defocus optical parameters to construct a physically constrained scattering PSF, we perform a \textit{forward reconstruction} of the instantaneous non-uniform scattering background via convolution operations. Subsequently, observations of polar coronal hole regions are introduced as physical constraints to determine the absolute intensity scaling coefficient of the background model, thereby achieving high-fidelity subtraction of the additive stray light component.

Crucially, to verify whether the correction algorithm preserves the intrinsic coronal signal, we established a ``ground--space'' cross-validation framework. Using simultaneous EUV observations from SDO/AIA as a morphological benchmark, we confirmed that while effectively stripping away high-intensity additive stray light, this method accurately restores the physical morphology of fine structures, such as streamers, that were previously obscured. This strategy, combining ``real-time physical monitoring'' with ``spatial data validation,'' achieves a leap from mere ``image enhancement'' to rigorous \textit{``physical restoration,''} providing a reliable solution for acquiring high-photometric-precision scientific data with ground-based coronagraphs.

The remainder of this paper is organized as follows: Section \ref{sec:instrument} details the instrument principles of the SICG, the dual-path monitoring system, and the data pre-processing workflow. Section \ref{sec:method} systematically elucidates the forward modeling method based on the Gaussian scattering kernel, including the optimization inversion of parameter $K$ and the fitting strategy for the intensity coefficient $C$. Section \ref{sec:results} presents the correction results under various contamination conditions and verifies the algorithm's robustness and scientific validity through statistical analysis (RMS, Signal-to-Background Ratio) and cross-comparison with SDO/AIA space-based data. Finally, Section \ref{sec:discussion} provides a discussion of the results and concludes the paper.

\section{Instrument and Data}
\label{sec:instrument}

\subsection{Instrument Overview}
\label{subsec:instrument}

\begin{figure}[ht!]
    \centering
   
    \includegraphics[width=0.95\textwidth]{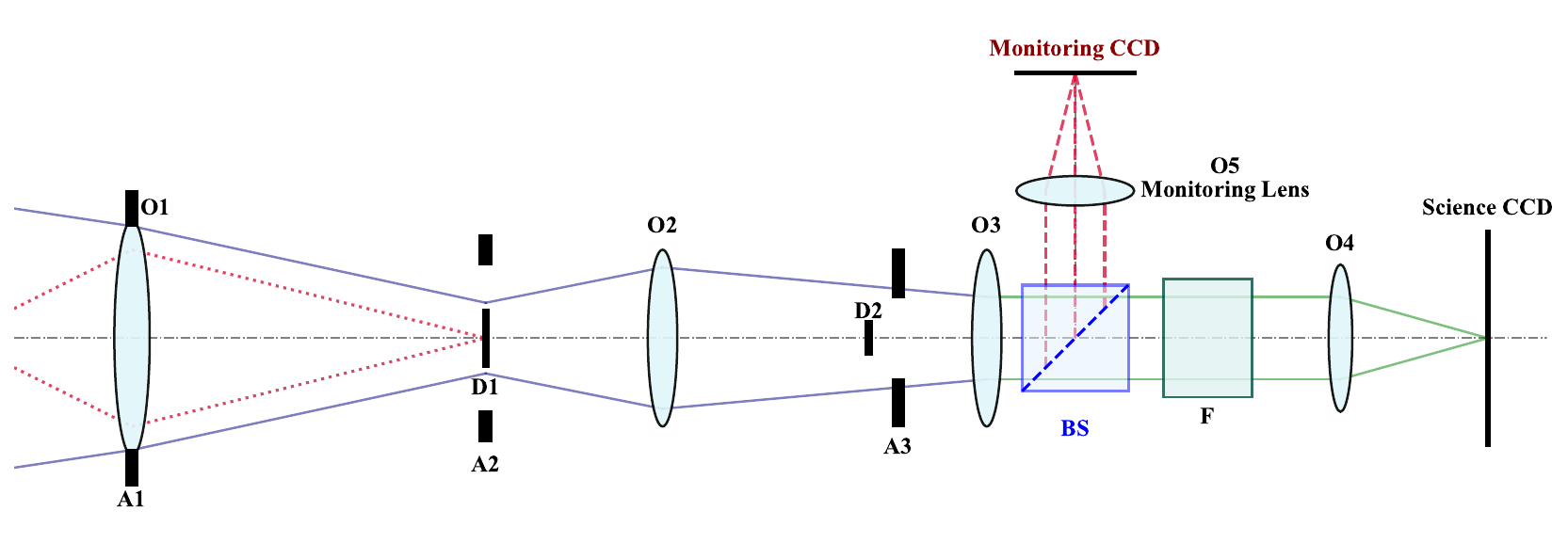}
    
    \caption{Schematic diagram of the optical layout of the SICG. The blue solid lines represent the coronal light path, while the red dotted lines indicate direct sunlight suppressed by the internal occulter (D1). Instrumental stray light is further mitigated by the Lyot stop (A3) and Lyot spot (D2). Key optical components are labeled as follows: O1, objective lens; A1, aperture stop; A2, field stop; O2, field lens; O3, relay lens; F, tunable filter; and O4, imaging lens;O5, monitoring lens. The monitoring channel relies on a precise pupil relay (O1 $\rightarrow$ A3 $\rightarrow$ Monitoring CCD) to sharply image the surface dust on the objective lens.}
    
    \label{fig:optical_path}
\end{figure}

The data analyzed in this study were obtained by the SICG, deployed at the Lijiang Yulong Observatory (elevation 3200 m) of Yunnan Observatories, a site characterized by excellent coronal observing conditions \citep{Tan2002}. The SICG is a ground-based internally occulted coronagraph developed for the second phase of the Chinese Meridian Project (CMP-II), designed for high-resolution imaging of the inner corona within a field of view (FOV) of $1.05$--$2.0 \, R_\odot$. Its primary observing passbands cover the coronal green line (\ion{Fe}{14} 530.3 nm) and red line (\ion{Fe}{10} 637.4 nm). The main optical system features a 200 mm aperture objective lens (O1, see Figure \ref{fig:optical_path}) with an effective focal length of approximately 2000 mm. Detailed descriptions of the instrument design specifications and performance parameters can be found in \citet{Huang2026} and \citet{Liu2025-SICG}.The observational data from the SICG, including the datasets utilized in this research, are publicly available and can be accessed through the Chinese Meridian Project portal at the National Space Science Data Center (\url{https://www.nssdc.ac.cn/nssdc_zh/html/siteinfo_with.html?6}).
As illustrated in Figure \ref{fig:optical_path}, the SICG incorporates rigorous multi-stage stray light suppression measures. The intense direct sunlight focused by the objective lens is intercepted by an internal occulter (D1) located at the primary focal plane; the angular diameter of D1 is designed to be 1.05 times that of the solar disk. To suppress diffracted light from the edge of the aperture stop (A1), a Lyot stop (A3) is employed in the optical path. Furthermore, ghost images generated by internal reflections within the objective lens are re-imaged by the field lens (O2) onto a Lyot spot (D2) for elimination. A detailed evaluation of the optical system design and the suppression level of static instrumental stray light is provided by \citet{Huang2026}. Notwithstanding these suppression mechanisms, as the primary optical element directly exposed to the full solar flux, the objective lens surface inevitably accumulates atmospheric dust. Since these dust particles are located at the instrument's entrance pupil, the resulting scattered light cannot be effectively intercepted by the downstream Lyot stop, thereby becoming the dominant source of residual stray light.

\subsection{Dual-Path Real-time Monitoring System}
\label{subsec:monitoring}

To quantify the dust contamination on the objective lens in real-time, the SICG incorporates a unique dual-path monitoring system \citep{Liu2025a}. As illustrated in the unified optical layout in Figure 1, a beam-splitting prism (BS) is positioned in the optical path immediately before the tunable filter (F) to divide the incident light into two distinct channels:

\begin{enumerate}
    \item The Science Channel (Transmitted): The transmitted beam continues through the main imaging lens (O4) and is focused at infinity onto the Science CCD, capturing the narrowband spectral images of the inner corona.
    
    \item The Monitoring Channel (Reflected): The reflected beam is directed into an auxiliary monitoring system (further detailed in Figure 2). Crucially, the imaging of the surface dust in this channel relies on a strict \textit{pupil-conjugate} optical configuration rather than conventional image-plane conjugation. Specifically, the objective lens (O1) serves as the entrance pupil of the system. The field lens (O2) re-images this entrance pupil onto the Lyot stop (A3), creating a secondary pupil plane. Within the monitoring channel, a dedicated monitoring lens group (O5) functions as a pupil-relay optic to project the A3 plane onto the Monitoring CCD. 
\end{enumerate}

Through this precise pupil relay chain (O1 $\rightarrow$ A3 $\rightarrow$ Monitoring CCD), the focal plane of the monitoring channel is optically conjugate to the front surface of the objective lens. This configuration uniquely enables the system to capture sharply resolved, high-resolution spatial distributions of the objective dust particles. Conversely, because the coronal signal originates from infinity, it remains entirely out of focus in this pupil-conjugate channel, diffusing into a uniform background that does not interfere with dust extraction. Furthermore, this beam-splitting design achieves strict temporal synchronization between the scientific observations and the instrument status monitoring, providing decisive, real-time prior information for the forward-modeling-based stray light correction described in Section 3.

\begin{figure}[ht!]
    \centering
    \includegraphics[width=0.85\textwidth]{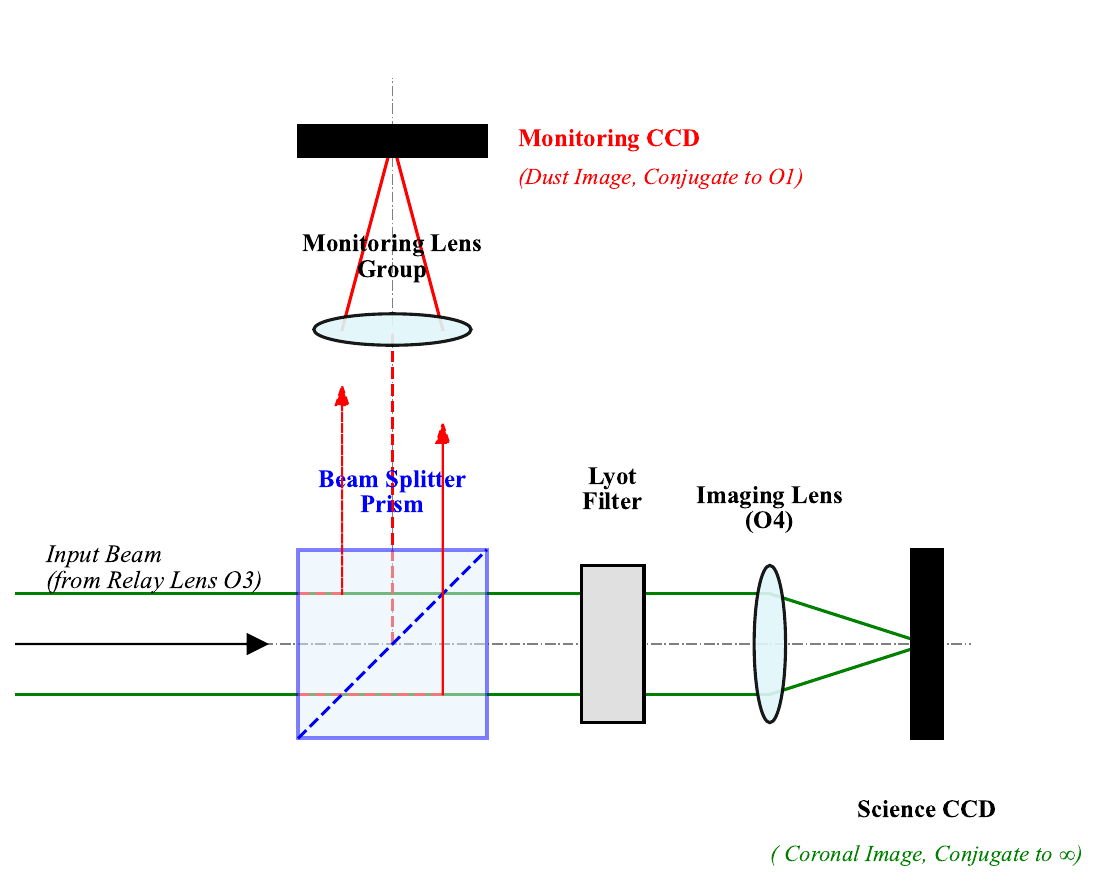}
    
    \caption{Schematic diagram of the dual-path monitoring system. A beam-splitting prism divides the incident light into two channels: the science channel (green path), which is focused at infinity for coronal imaging, and the monitoring channel (red path), which is focused on the objective lens surface for dust monitoring. This configuration allows for the simultaneous acquisition of scientific data and the dust source map.}
    
    \label{fig:Fig2}
\end{figure}

\subsection{Data Acquisition and Pre-processing} \label{subsec:data_processing}

To systematically evaluate the efficacy and robustness of the scattering correction algorithm under varying instrument conditions, we selected three representative SICG datasets from the 2024 observation logs. These datasets correspond to different contamination levels of the objective lens surface: low contamination (Clean, Oct 31), moderate contamination  (Moderate, Nov 20), and high contamination (Dirty, Dec 12). All observations targeted the inner coronal emission line of \ion{Fe}{14} 530.3 nm. For each dataset, the instrument operated in a spectral scanning mode, acquiring images at five wavelength positions covering the line center and wings, while the monitoring channel simultaneously captured the corresponding images of the objective lens surface.

During routine observations, the dual-path system of the SICG simultaneously generates two independent data streams, which are acquired by two distinct cameras:
\begin{enumerate}
    \item \textit{Coronal Science Data}: Captured by the \textit{Science CCD} in the transmitted optical path. These images record the true coronal radiation signals but are severely contaminated by instrumental stray light induced by dust.
    \item \textit{Dust Source Data}: Captured by the \textit{Monitoring CCD} in the reflected optical path. These images precisely record the spatial distribution of dust particles on the objective lens surface at the exact instant the corresponding science images are acquired.
\end{enumerate}

Prior to the application of the scattering model proposed in this paper, the raw observation data from the science camera underwent standard instrumental calibration procedures. Let $I_{raw}(x,y)$ denote the raw observation image, $I_{dark}(x,y)$ the dark current image (to avoid symbol confusion with the dust source function introduced later, the previous notation $D$ is discarded here), and $I_{flat}(x,y)$ the normalized flat-field image. The pre-processed image intensity, $I_{pre}(x,y)$, is obtained via the following standard formula:

\begin{equation}
    I_{pre}(x,y) = \frac{I_{raw}(x,y) - I_{dark}(x,y)}{I_{flat}(x,y) - I_{dark}(x,y)} \times \langle I_{flat} \rangle
    \label{eq:flat_field}
\end{equation}

where $\langle I_{flat} \rangle$ represents the mean value of the flat-field image, used to maintain the data scale. It is crucial to point out that the flat-field correction represented by Equation (\ref{eq:flat_field}) essentially addresses multiplicative gain errors. However, the stray light caused by dust on the objective lens physically manifests as an additive term superimposed on the coronal signal. Consequently, the flat-field division operation cannot effectively remove it.

Following flat-field correction, two critical steps of geometric registration are required. The \textit{registration problem} fundamentally arises because the science camera and the monitoring camera reside in two independent physical optical branches. Inherent mechanical installation tolerances of the beam-splitting prism and lens groups inevitably lead to translational and minor rotational offsets between their respective fields of view. If left uncorrected, the scattering background generated based on the monitoring camera would fail to align accurately with the science image in spatial coordinates.

To resolve this, we implemented the following registration methods:

The first step is \textit{absolute geometric calibration}. Based on the verification of structural similarity between \ion{Fe}{14} and EUV 211 \AA\ by \citet{Zhang2022_Comparison}, we utilized simultaneous SDO/AIA 211 \AA\ observation data to align the SICG images via cross-correlation. Adopting the high-precision image registration and coordinate mapping methods proposed by \citet{Sha2025C}, we established a precise heliocentric coordinate system, thereby defining accurate polar regions for the subsequent background fitting.

The second step is the \textit{relative registration} between the two cameras. To correct the aforementioned inherent mechanical offsets between the monitoring and science optical paths, we utilized the physical edge of the internal occulter, which is clearly imaged in the fields of view of both cameras, as a common fiducial feature. In practice, we first extracted local regions of interest (ROIs) containing the occulter edge from both images. Subsequently, a phase cross-correlation algorithm was employed in the frequency domain to calculate the sub-pixel translational and minor rotational offsets of the monitoring image relative to the science image. Finally, bicubic interpolation was used for spatial resampling and geometric transformation of the monitoring image, mapping it precisely onto the coordinate system of the science image.

The pre-processed science data resulting from the flat-field correction and absolute registration are defined as \textit{Level 1 data} ($I_{L1}$) in this study. These data, together with the aligned dust source maps, serve as the input for the forward modeling algorithm described in Section \ref{sec:method}.

\section{Forward Modeling of the Dust Scattering Background} \label{sec:method}

The core philosophy of this study utilizes the prior information provided by the dual-path monitoring system to reconstruct the stray light background induced by dust on the objective lens through \textit{forward physical modeling}. This model treats the observed stray light field as the convolution of discrete dust point sources with the instrument's defocus PSF. The modeling workflow consists primarily of three steps: constructing the dust source function, defining the scattering kernel function, and retrieving the model parameters ($K$ and $C$) through observational data inversion.

\subsection{Overview of the Data Processing Pipeline}
\label{sec:overview}

To clarify the position of the proposed dust correction algorithm within the overall data processing framework of the SICG, Figure 3 presents a comprehensive flowchart of the correction pipeline. 

Because the dust scattering induced by the objective lens physically manifests as a severe additive and non-uniform background, its correction must be strategically positioned after the basic multiplicative calibrations but prior to any general background subtractions. As illustrated in the flowchart, the scientific raw data ($I_{raw}$) first undergo standard dark and flat-field corrections ($I_{pre}$). This step is followed by absolute geometric registration with SDO/AIA to yield the Level 1 data ($I_{L1}$). Meanwhile, the simultaneously acquired monitoring raw images undergo high-pass filtering and masking to extract the dust spatial distribution. To eliminate the intrinsic mechanical offset between the two optical paths, a relative geometric registration is performed. Using the physical edge of the internal occulter as a common reference feature, we align and map the extracted dust features onto the science image coordinate system to generate the final dust source map ($D(x, y)$).

The core of our method is the forward modeling loop, which is indicated by the dashed box in Figure 3. In this loop, the dust source map is convolved with a generated Gaussian kernel to construct the normalized scattering basis ($M_{base}$). An optimization process is then performed to determine the optimal characteristic width ($K$) and fit the intensity scaling coefficient ($C$) using the polar regions as a reference. Finally, the optimized background model ($B_{scat}$) is subtracted from the Level 1 data to produce the structurally clean, corrected image ($I_{corr}$). 

Only after this non-uniform morphological distortion is thoroughly stripped away can the subsequent standard procedures be physically justified and accurately executed in future scientific applications. These procedures primarily include uniform sky background subtraction and absolute radiometric calibration. The specific mathematical and physical details of this forward modeling loop are systematically elucidated in the following subsections.

\begin{figure}[htbp]
    \centering
    \includegraphics[width=\textwidth]{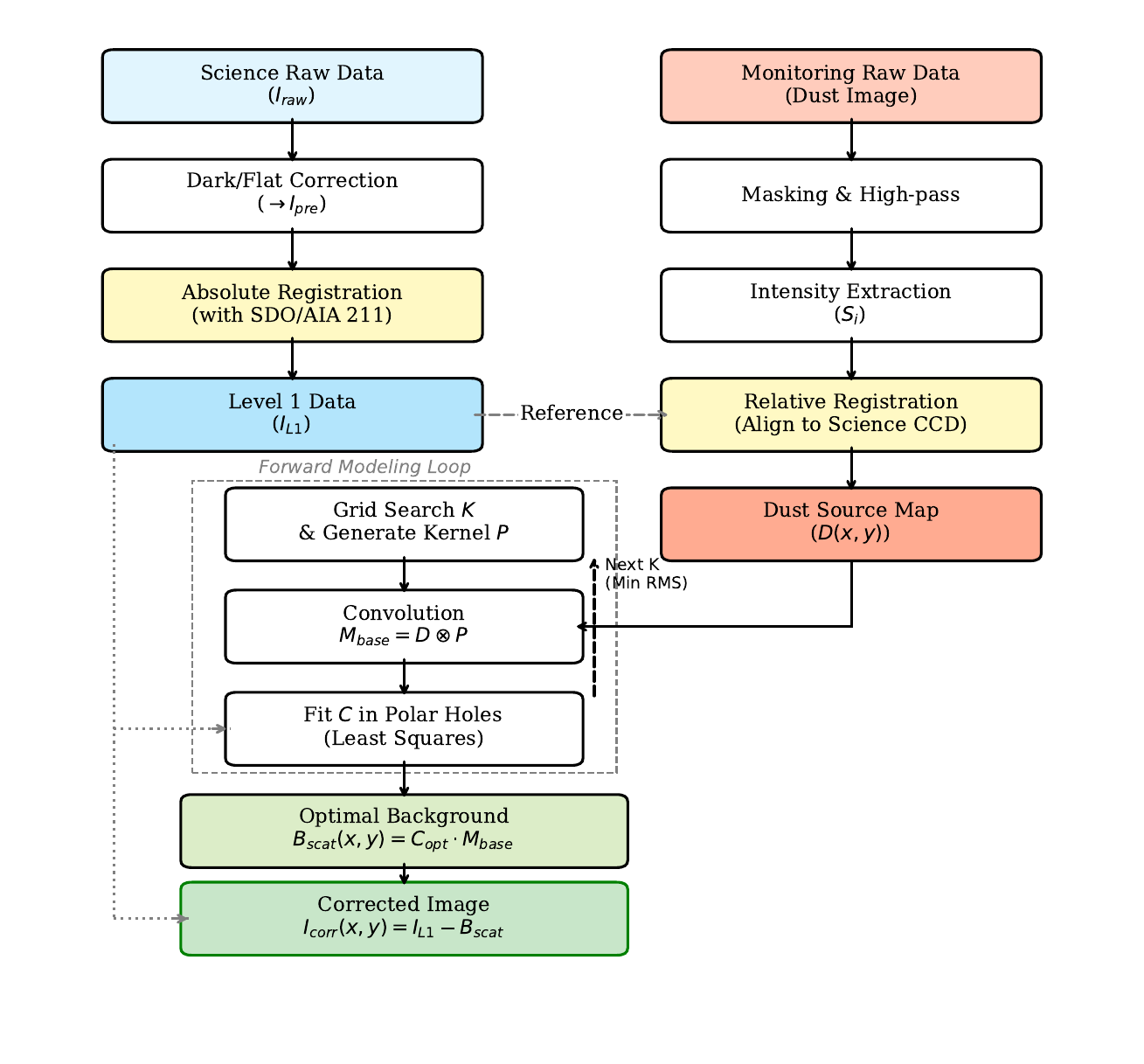}
     \caption{Flowchart of the comprehensive data processing pipeline, illustrating the strategic placement of the forward-modeling dust correction loop between the basic calibrations (Level 1) and the final scientific data products.}
     \label{fig:flowchart}
 \end{figure}
 
\subsection{Construction of the Dust Source Function} \label{subsec:source_function}

\begin{figure*}[ht!]
    \centering
    \includegraphics[width=\textwidth]{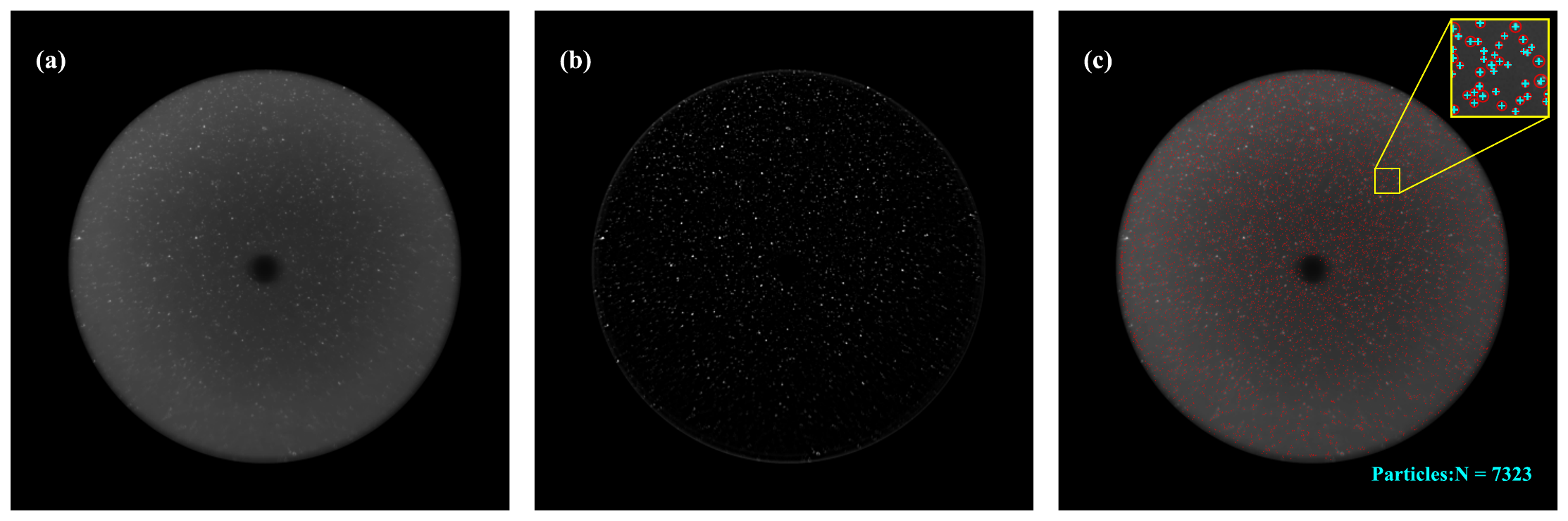}
    \caption{Process of extracting objective lens dust parameters. 
\textbf{(a)} The original monitoring image, showing the dust distribution superimposed on a non-uniform illumination gradient background. 
\textbf{(b)} The high-pass enhanced image, where the low-frequency background is suppressed, isolating the high-frequency signal components representing the dust particles. 
\textbf{(c)} Final detection results overlaid on the original image. Red dots indicate the centroids of all identified dust particles. The zoomed-in inset (top right) illustrates the precision of the detection, where cyan crosses mark the centroids and red circles delineate the equivalent boundaries of individual particles. The integrated intensity within these regions is utilized to construct the weighted source function $D(x,y)$.}
    \label{fig:Fig3a}
    
\end{figure*}

The primary task in modeling the stray light is to accurately extract the spatial distribution and relative scattering intensity of dust particles from the monitoring channel images. As shown in Figure \ref{fig:Fig3a}(a), the raw monitoring image clearly reveals the dust on the objective lens but is superimposed with a low-frequency background gradient caused by the illumination system and stray reflections from the lens mount. To mitigate these artifacts, we first applied a circular geometric mask based on the registration parameters described in Section \ref{subsec:data_processing} to zero out regions outside the effective clear aperture. Subsequently, to suppress the non-uniform illumination within the field of view, a \textit{high-pass filter} was applied to the image. As demonstrated in Figure \ref{fig:Fig3a}(b), this process effectively removed the slowly varying low-frequency background, isolating the high-frequency signal components representing the dust particles. Based on the enhanced image, we calculated the standard deviation of pixel intensities in dust-free regions, denoted as $\sigma_{bg}$, to characterize the background noise level. Using an adaptive thresholding algorithm, connected regions with intensities exceeding $3\sigma_{bg}$ were identified as valid dust scattering sources (Figure \ref{fig:Fig3a}(c)).

Following feature identification, to account for the physical differences in scattering capacity due to dust size and opacity, we constructed a weighted source function rather than using a simple binary mask. For each identified dust region, the \textit{integrated intensity} from the original image was extracted and assigned as the weight $S_i$. The final source function $D(x,y)$ is defined as:
\begin{equation}
    D(x,y) = \sum_{i} S_i \cdot \delta(x-x_i, y-y_i)
    \label{eq:source_func}
\end{equation}
where $(x_i, y_i)$ represents the centroid coordinates of the $i$-th dust particle. As correctly pointed out by the fundamental scattering principles, the weight $S_i$ is strictly proportional to the total scattered intensity produced by the dust grain with index $i$, reflecting its relative scattering cross-section. Consequently, this source function $D(x,y)$ serves as an image array of the same dimensions as the science image, which is mostly zero except at the discrete locations of the identified dust particles. $D(x,y)$ constitutes the core input for the subsequent convolution operations.

\subsection{Formulation of the Scattering Point Spread Function} \label{subsec:psf_formulation}

Physically, the scattering of incident sunlight by dust particles on the objective lens surface is governed by Mie scattering theory \citep{VandeHulst1957}. In the operational context of ground-based coronagraphs, the typical size $d$ of dust particles settling on the lens surface generally ranges from $10^0$ to $10^2$ $\mu\text{m}$, significantly exceeding the observing wavelength of the SICG ($\lambda=530.3$ nm). According to Mie theory, when the size parameter $x = \pi d / \lambda \gg 1$, the scattering phase function exhibits strong forward scattering characteristics. This implies that the vast majority of scattered energy is not distributed isotropically as in Rayleigh scattering, but is instead highly concentrated within a narrow angular range along the line of sight, forming a prominent diffraction main lobe \citep{Fineschi1993}. This intense forward-scattered beam can propagate through the subsequent stop system and enter the field of view, constituting the physical origin of the high-intensity stray light background in the observed images.

Although the theoretical Mie scattering pattern of a single ideal spherical particle contains complex concentric diffraction rings, under the actual observational configuration of the SICG, these high-frequency coherent features undergo significant smoothing, eventually degenerating into a smooth veiling glare. This physical approximation is primarily attributed to the instrument's extreme defocus effect: since the dust is located on the front surface of the objective lens, the imaging system operates in a regime of massive defocus ($\Delta z \approx f$) relative to the science focal plane focused at infinity, causing significant geometric spreading of the scattered beam on the image plane. Furthermore, the finite observing bandwidth ($1.2$ \AA) induces phase decoherence in high-order diffraction structures, while the statistical averaging effect arising from the randomness in size, shape, and refractive index of naturally settled dust further eliminates specific angular diffraction features. Given these multiple physical mechanisms, and consistent with classical models of scattering from optical surface contamination \citep{Harvey2012, DeForest2009}, the Gaussian function is widely regarded as a valid and standard approximation for describing the PSF of such defocused surface scattering.

Based on the aforementioned physical approximations, we define the instrument's scattering kernel function $P(x, y)$ as a normalized two-dimensional Gaussian distribution. This function characterizes the spatial morphology formed by a unit-intensity point source on the focal plane due to defocusing:

\begin{equation}
    P(x, y; K) = \frac{1}{\pi K^2} \exp\left( - \frac{x^2 + y^2}{K^2} \right)
    \label{eq:psf_kernel}
\end{equation}

This model incorporates two key physical parameters. The \textit{characteristic width} ($K$) defines the spatial extent of the scattering halo ($1/e$ decay radius). Since the defocus scale of the instrument (millimeters) is orders of magnitude larger than the physical size of the dust particles (micrometers), the width of the PSF is dominated by the instrument's optical geometry rather than individual particle sizes. Consequently, $K$ is treated as a uniform \textit{effective instrument parameter}, the optimal value of which will be determined in Section \ref{subsec:optimization_k}. The other parameter, the \textit{amplitude weight} ($S_i$), is embedded within the dust source function $D(x, y)$ constructed in Section \ref{subsec:source_function}. Physically, it represents the relative scattering cross-section, ensuring that particles with larger geometric sizes contribute a proportionally increased scattering flux.

Given the linear superposition nature of incoherent stray light, the total scattering background $B_{scat}(x, y)$ is generated by the convolution of the source function $D(x, y)$ with the scattering kernel $P(x, y; K)$. Since $D(x,y)$ is an image array that is mostly zero except at the discrete locations of the dust particles, this convolution can also be intuitively represented as a sum over its non-zero elements, scaled by a global coefficient $C$:

\begin{equation}
    B_{scat}(x, y) = C \cdot [D(x, y) \otimes P(x, y; K)] = C \cdot \sum_{i=1}^{N} S_i \cdot P(x - x_i, y - y_i; K)
    \label{eq:convolution}
\end{equation}

where the symbol $\otimes$ denotes the convolution operation, and $N$ is the total number of identified dust particles. The global coefficient $C$ serves as a photometric calibration factor to match the flux scales between the monitoring and science channels, which will be fitted using observational constraints in Section \ref{subsec:fitting_c}.

\subsection{Optimization of the Characteristic Width Parameter K} \label{subsec:optimization_k}

The characteristic width $K$ of the scattering kernel is the critical parameter determining the spatial morphology of the background model. As discussed in Section \ref{subsec:psf_formulation}, while geometric optics provides an upper limit for the defocus blur radius ($R_{geo} \approx 10 \, R_\odot$ for SICG), the effective scattering width is significantly smaller than this geometric limit due to the pronounced forward scattering characteristics of micron-sized dust particles. Considering the randomness in particle size, shape, and refractive index of naturally settled dust, a rigorous analytical derivation of $K$ is practically infeasible. Therefore, this study adopts a data-driven optimization strategy to retrieve the optimal $K$ value.

We selected the ``High Contamination'' observation data, which exhibits the highest signal-to-noise ratio, as the calibration benchmark to ensure maximum sensitivity of the optimization process to morphological variations in dust artifacts. The optimization procedure was conducted as follows: a series of trial values were defined within the parameter space $K \in [0.1, 1.0] \, R_\odot$; for each $K$, the corresponding scattering basis was generated using Equation (\ref{eq:convolution}), and the global intensity coefficient $C$ was fitted. Subsequently, the Root Mean Square (RMS) of the residuals within the polar coronal hole region ($1.1$--$1.8 \, R_\odot$) was calculated as the cost function to be minimized.

\begin{figure}[ht!]
    \centering
    \includegraphics[width=0.85\textwidth]{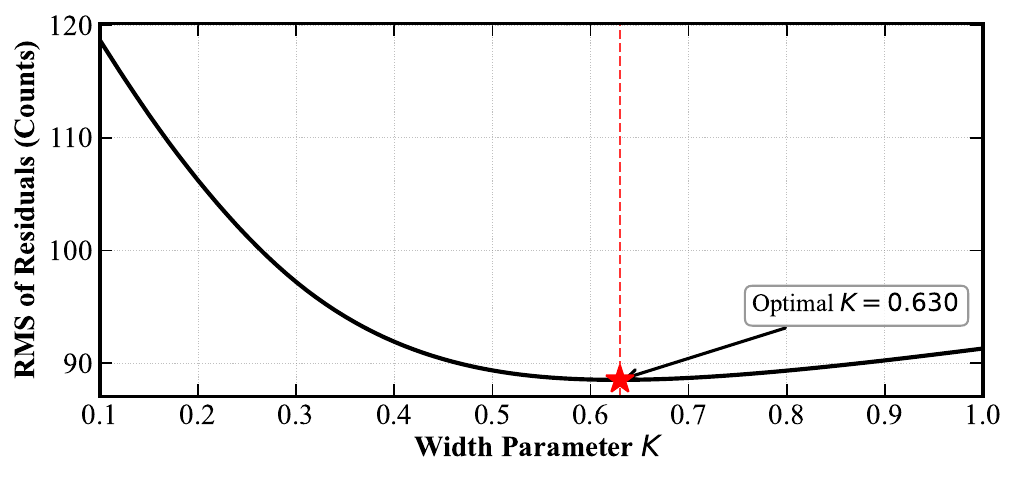}
    
    \caption{Optimization of the scattering width parameter $K$. The solid curve represents the Root Mean Square (RMS) of the residuals in the polar coronal hole region as a function of the trial width $K$. A clear global minimum is identified at $K = 0.630 \, R_\odot$ (marked by the red star), indicating the optimal morphological match between the scattering model and the observed dust artifacts.}
    
    \label{fig:Fig4}
\end{figure}

Figure \ref{fig:Fig4} illustrates the variation of the residual RMS as a function of the parameter $K$. The curve exhibits a distinct concave structure, indicating the existence of a unique global minimum. The behavior of the RMS curve can be explained by the geometric properties of the convolution. Specifically, in the underestimation regime ($K < 0.5$), the kernel width is smaller than the actual dust halo, making the model profile spatially too narrow. Although the central peak may match, the extended wings of the dust profile are not fully covered by the model, resulting in ring-like positive residuals in the corrected image. Conversely, in the overestimation regime ($K > 0.8$), the model spot becomes overly broad and smooth. To match the total flux during the least-squares fitting, the model inevitably lowers the central peak and spills over outward, causing negative dips at the dust centers while eroding the surrounding true background. Only when $K$ matches the true physical dispersion does the spatial variance of the residuals reach a minimum.

These results indicate that a characteristic width of $0.630 \, R_\odot$ yields the best match between the constructed physical model and the observed dust scattering morphology. Consequently, we fix $K$ at $0.630 \, R_\odot$ and apply it as a constant physical parameter for the stray light correction of all observational data in this instrument state.

\subsection{Calibration of the Global Intensity Coefficient $C$} \label{subsec:fitting_c}
Following the determination of the characteristic width $K$, the final step in constructing the scattering background model is to solve for the global intensity scaling coefficient $C$.  Based on the convolution model described in Equation (4), we isolate the spatial morphology component from the absolute intensity scalar by defining a normalized scattering basis, $M_{base}(x, y)$:

\begin{equation}
M_{base}(x, y) = D(x, y) \otimes P(x, y; K)
\label{eq:mbase}
\end{equation}

This basis $M_{base}(x, y)$ completely captures the complex spatial distribution and relative intensity differences of the dust artifacts within the field of view, generated by convolving the monitored dust source function with the Gaussian kernel. Subsequently, a global linear coefficient $C$ is required to cross-calibrate the numerical magnitude of this basis to the radiometric scale of the science observations. This is necessary because the monitoring and science cameras reside in different optical branches and operate with distinct optical efficiencies, exposure times, and electronic gains. Consequently, the final scattering background model is expressed as the product of $M_{base}$ and $C$.

To constrain and solve for the coefficient $C$, we introduce a calibration region with clear physical justification. Specifically, we selected the polar coronal holes as the calibration region $\mathcal{R}_{pole}$, defined by a radial range of $1.1$--$1.8 \, R_\odot$ and restricted to latitudes within $\pm 10^\circ$ of the solar poles. During periods of low solar activity, this region is characterized by significantly reduced electron density, resulting in extremely weak K-corona emission. The contribution of the intrinsic coronal radiance to the observed brightness is therefore negligible compared to the stray light background. Consequently, in this region, the observed brightness is dominated by instrumental stray light, justifying the approximation $I_{obs}(x,y) \approx B_{scat}(x,y)$ \citep{vandeHulst1950, November1996, DeForest2009}.

Under this approximation, we employ the least-squares fitting method to quantitatively retrieve $C$. The sum of squared residuals within the polar coronal hole region is defined as the objective function:

\begin{equation}
    \chi^2 = \sum_{(x,y) \in \mathcal{R}_{pole}} \left[ I_{obs}(x,y) - C \cdot M_{base}(x,y) \right]^2
\end{equation}

By setting the partial derivative $\partial \chi^2 / \partial C = 0$, the analytical solution to this linear optimization problem is obtained as:

\begin{equation}
    C = \frac{\sum_{(x,y) \in \mathcal{R}_{pole}} I_{obs}(x,y) \cdot M_{base}(x,y)}{\sum_{(x,y) \in \mathcal{R}_{pole}} \left[ M_{base}(x,y) \right]^2}
    \label{eq:solution_c}
\end{equation}

The coefficient $C$ thus obtained achieves absolute intensity calibration between the stray light model and the scientific observation images. Subtracting the constructed scattering background $B_{scat}$ from the raw observational images significantly suppresses the impact of instrumental stray light on coronal observations, while preserving the spatial structure and physical information of the true coronal radiation with high fidelity.

It is crucial to emphasize that the least-squares optimization defined in Equation (5) is designed to minimize the spatial variance of the residuals within the polar regions, rather than forcing their absolute intensity values to zero. Because the forward model strictly reconstructs only the non-uniform instrumental stray light induced by objective dust, the subtraction process ($I_{corr} = I_{obs} - B_{scat}$) safely preserves the uniform, physically meaningful background components, such as atmospheric sky brightness, F-corona, detector readout noise, and faint intrinsic K-corona emissions (e.g., polar plumes). Therefore, the corrected polar regions retain a physically valid, non-zero intensity level, ensuring the photometric reliability of the data for subsequent studies of faint coronal structures.

\section{Results and Validation} \label{sec:results}

\subsection{Image Correction Performance and Physical Feature Recovery} \label{subsec:correction_performance}

\begin{figure*}[ht!]
    \centering
    \includegraphics[width=\textwidth]{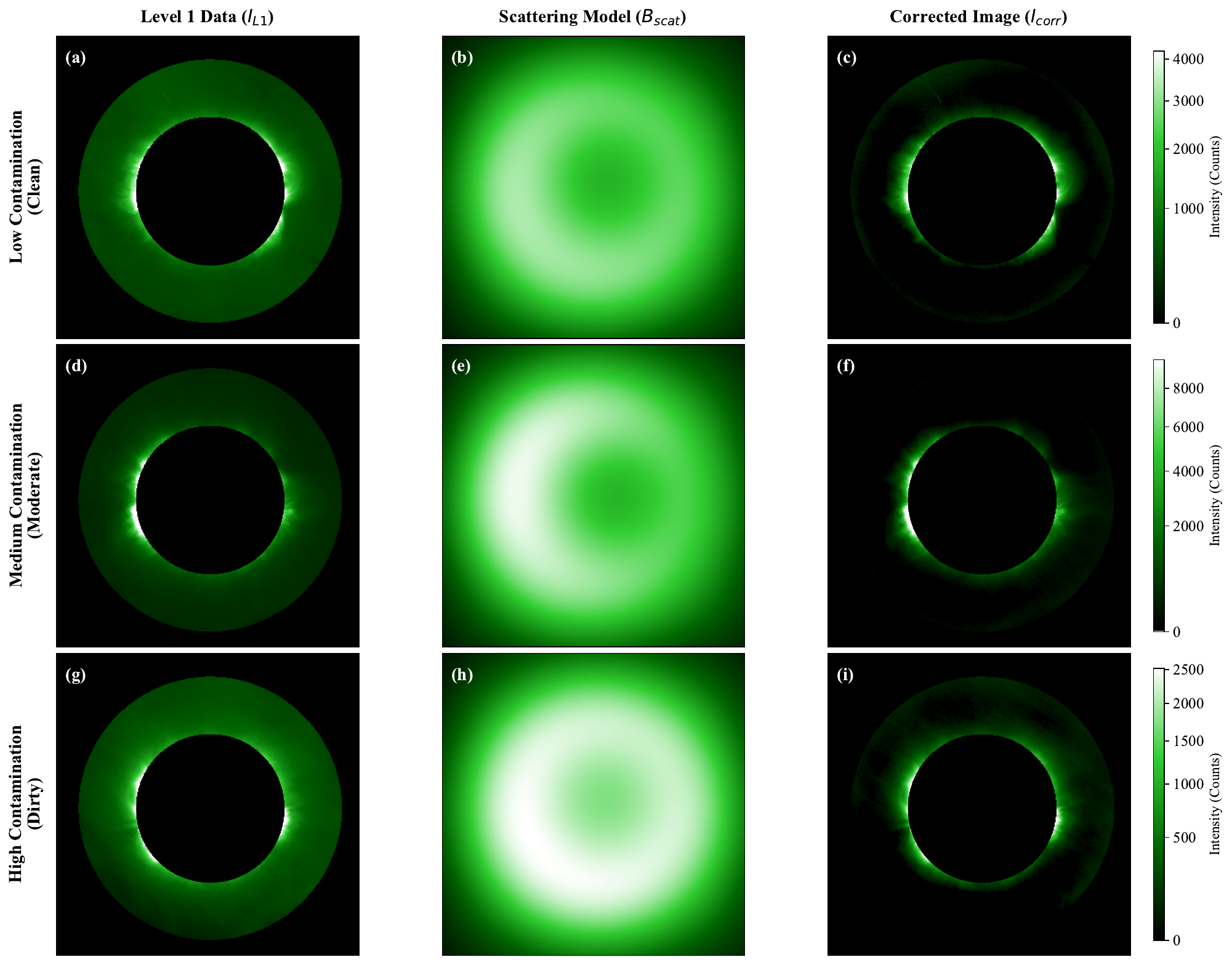}
    
    \caption{Visual comparison of the stray light correction performance under different objective lens contamination levels. 
    \textbf{Left column} (a, d, g): The Level 1 observed images ($I_{L1}$) after flat-field correction, corresponding to three representative states: Low (Clean), Medium (Moderate), and High (Dirty) contamination. 
    \textbf{Middle column} (b, e, h): The scattering background models ($B_{scat}$) constructed based on simultaneously acquired dust distribution and physical scattering modeling. 
    \textbf{Right column} (c, f, i): The corrected results ($I_{corr}$) after subtracting the stray light background. 
    }
    
    \label{fig:Fig5}
\end{figure*}

Figure \ref{fig:Fig5} visually illustrates the efficacy of the scattering correction algorithm across varying contamination levels. By comparing the pre-processed data (left column) with the corrected images (right column), it is evident that the diffuse stray light haze obscuring the field of view has been effectively removed. This is particularly noticeable for the high-intensity glare on the left side of the image, which originates from the asymmetric distribution of dust.

In the ``Dirty'' case (Figures \ref{fig:Fig5}(g)--(i), where scattering is most severe, the algorithm successfully disentangles the nearly overwhelmed coronal signals from the intense background, significantly enhancing the visibility of streamer structures. The scattering models displayed in the middle column accurately reproduce the non-uniform spatial morphology and radial distribution characteristics of the stray light under all three contamination conditions, demonstrating the model's excellent adaptability to variations in the contamination state.

These results indicate that the proposed method not only effectively eliminates the large-scale additive stray light background but also maintains consistent correction performance across different contamination regimes. By maximizing the preservation of the intrinsic spatial structure of coronal radiation, these findings validate the effectiveness of the forward modeling strategy based on monitoring data.

To validate the physical fidelity of the corrected data, we extracted radial intensity profiles from the equatorial streamer region (Position Angle $0^\circ \pm 10^\circ$) and performed fitting analysis. Considering that the inner coronal plasma is in a state of \textit{hydrostatic equilibrium}, its electron density and the resulting emission intensity theoretically follow an exponential decay law with radial distance \citep[e.g.,][]{vandeHulst1950, Aschwanden2005}. Consequently, we adopted the following exponential model to fit the observational data within the range of $1.05$--$1.65 \, R_\odot$:

\begin{equation}
    I(r) = A \cdot \exp[b(r-1)] + c
    \label{eq:exp_fit}
\end{equation}

where $c$ represents the residual additive background pedestal.

\begin{figure}[ht!]
    \centering
    \includegraphics[width=0.95\columnwidth]{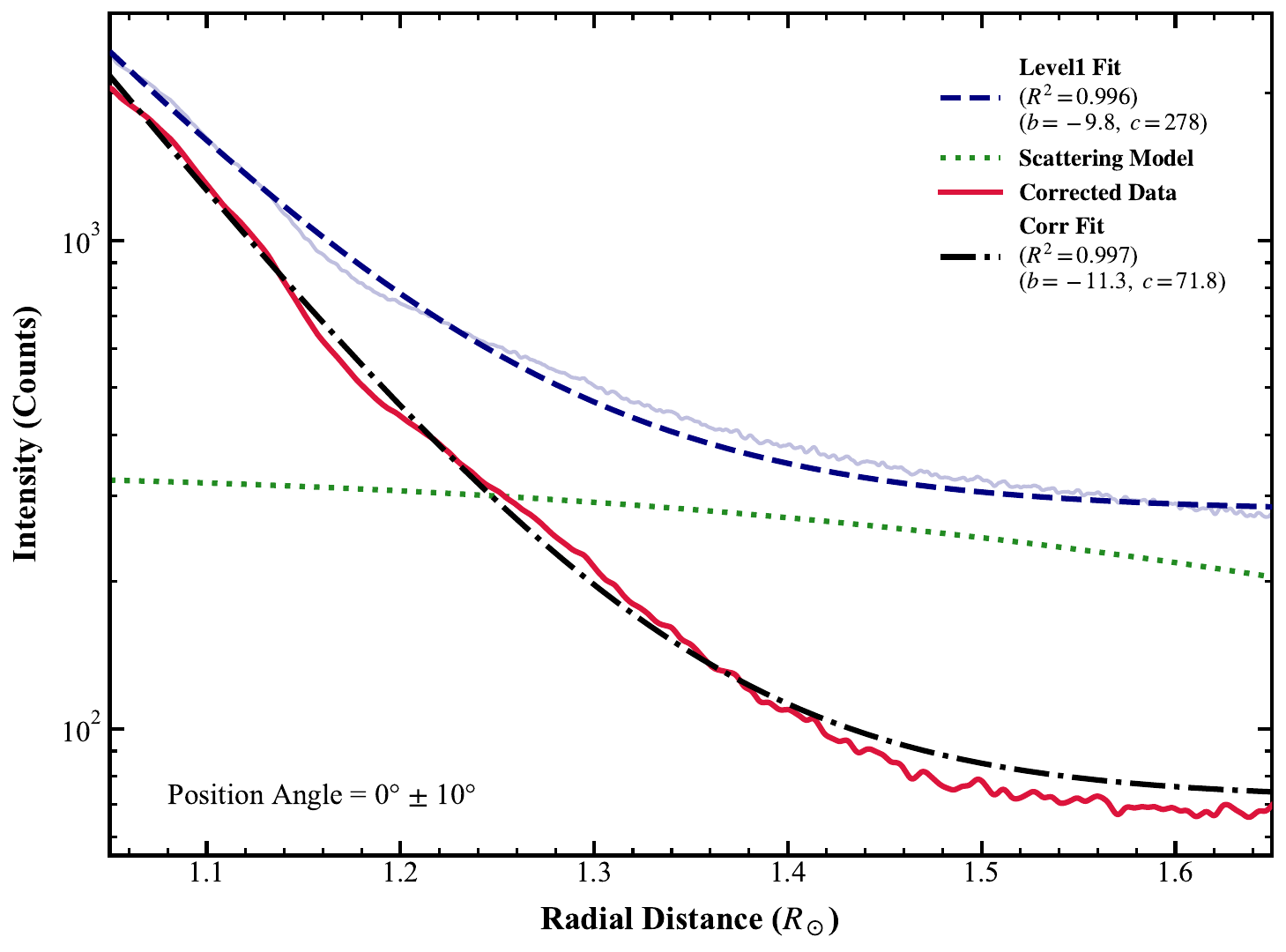}
    
    \caption{Radial intensity profiles and exponential fitting analysis in the equatorial streamer region (Position Angle $0^\circ \pm 5^\circ$). 
    The \textbf{light blue solid line} represents the flat-fielded Level 1 data, overlaid with its best-fit curve (\textbf{dark blue dashed line}). 
    The \textbf{green dotted line} shows the derived scattering background model ($B_{scat}$). 
    The \textbf{red solid line} denotes the dust-removed Level 2 data, which exhibits excellent agreement with the theoretical exponential model (\textbf{black dash-dot line}, $R^2 > 0.99$). 
    Note the significant reduction in the background constant $c$ (from 278 to 71.8) and the steepening of the decay index $|b|$ (from 9.8 to 11.3) after correction, indicating the effective removal of the additive stray light pedestal and the recovery of the intrinsic hydrostatic decay signature of the corona.}
    
    \label{fig:radial_profile}
\end{figure}

As shown in Figure \ref{fig:radial_profile}, the original Level 1 data (light blue solid line) exhibits a significant flattened tail beyond $1.3 \, R_\odot$. The exponential fitting results reveal a high background constant of $c \approx 278$ counts and a relatively small decay coefficient ($b = -9.8$). This indicates that diffuse additive stray light dominates the observational signal in the outer field of view, severely masking the intrinsic decay trend of the corona.

In contrast, the corrected data (red solid line) displays a steeper and cleaner decay profile, showing a high degree of agreement with the theoretical exponential model (black dash-dot line, $R^2 = 0.997$). Quantitative analysis further reveals the recovery of physical characteristics: the fitted background term $c$ drops significantly from 278 to 71.8, implying that approximately 75\% of the non-physical additive pedestal has been effectively stripped away. 

It is physically expected that the residual background term $c$ does not drop to identically zero in ground-based observations. This remaining stable pedestal of $\sim$71.8 counts represents the irreducible intrinsic atmospheric scattering (sky brightness), F-corona, and fundamental detector readout noise. Forcing this term to zero would unphysically over-subtract the natural sky background. Thus, the significant reduction of $c$ without reaching zero confirms that our algorithm successfully isolates and removes only the anomalous instrumental dust scattering, meticulously preserving the true photometric boundary conditions of the inner corona. Simultaneously, the magnitude of the decay coefficient $|b|$ increases from 9.8 to 11.3. This steeper gradient is consistent with the density distribution characteristics of the inner corona under hydrostatic equilibrium, corresponding to a more reasonable coronal scale height and temperature structure.

In summary, the algorithm not only visually eliminates artifacts but also photometrically restores the physical morphology of the coronal emission, significantly enhancing the signal-to-noise ratio and scientific utility of the data.

\subsection{Quantitative Evaluation and Robustness Analysis} 
\label{subsec:quantitative_eval}
To quantitatively evaluate the performance of the scattering correction algorithm, we conducted a detailed assessment of three observational datasets with varying degrees of contamination, based on the statistical metrics listed in Table \ref{tab:correction_stats}. The selected cases span a wide range from a low contamination state (534 particles) to a high contamination state (7323 particles), representing a difference in dust load of more than a factor of 14. In response to these vastly different inputs, the forward model fitted the corresponding intensity calibration coefficient $C$ via the least-squares method. The adaptive adjustment of this value with respect to the dust load demonstrates the algorithm's capability to automatically match the optimal subtraction intensity according to the actual total flux.
In terms of background noise suppression, statistical results from the polar coronal holes reveal that the original RMS noise in all test cases decreased from hundreds of counts to approximately 100 counts after correction, achieving an average reduction of about 67\%. This indicates that the majority of the non-physical scattering energy induced by dust has been effectively removed.

Furthermore, to quantitatively evaluate the distinguishability of coronal structures relative to the background, we introduce the Signal-to-Background Ratio (SBR) as a key evaluation metric. Specifically, we define the equatorial streamer region (Position Angle $0^\circ/180^\circ \pm 45^\circ$), where coronal emission is strongest, as the \textit{signal region} ($R_{sig}$), and the polar coronal hole region (Position Angle $90^\circ/270^\circ \pm 10^\circ$), where emission is weakest, as the \textit{background region} ($R_{bg}$). The SBR calculated within the inner coronal core region ($1.05$--$1.20 \, R_\odot$) directly reflects the prominence of coronal structures relative to the background pedestal.

Our analysis indicates that the corrected SBR achieves substantial improvement across all operating conditions. Particularly in the heavy contamination case, the SBR jumps from a practically unusable 1.1 to 3.9, representing an enhancement factor of up to 3.7$\times$. This result confirms that even under conditions of extreme contamination, the algorithm can effectively restore the scientific utility of the data.

\renewcommand{\arraystretch}{1.15} 

\begin{deluxetable*}{lccccc} 
\tabletypesize{\small} 

\tablecaption{\textbf{Quantitative assessment of the scattering correction performance.}\label{tab:correction_stats}}
\tablecolumns{6}
\tablewidth{0pt} %

\tablehead{
    \colhead{\textbf{Dataset}} & 
    \colhead{\textbf{Obs. Time}} & 
    \colhead{\textbf{Dust Load}} & 
    \colhead{\textbf{Fitted Coeff}} & 
    \colhead{\textbf{Background Noise}} & 
    \colhead{\textbf{Signal-to-Background}} \\
    \colhead{(Condition)} & 
    \colhead{(UT)} & 
    \colhead{($N_{part}$)} & 
    \colhead{($C$)} & 
    \colhead{$I_{L1} \to I_{corr}$ (RMS)} & 
    \colhead{$I_{L1} \to I_{corr}$ (SBR)}
}
\startdata
\textbf{Case 1} (Clean)    & 2024-10-31 & 534 & 0.161 & $483 \to 152$ (\textbf{68\%}) & $1.8 \to 4.3$ (\textbf{2.4$\times$}) \\
\textbf{Case 2} (Moderate) & 2024-11-20 & 1,148 & 0.123 & $560 \to 211$ (\textbf{62\%}) & $1.2 \to 3.9$ (\textbf{3.2$\times$}) \\
\textbf{Case 3} (Dirty)    & 2024-12-12 & 7,323 & 0.052 & $297 \to 88$~~ (\textbf{70\%}) & $1.1 \to 3.9$ (\textbf{3.7$\times$}) \\
\enddata

\tablecomments{The scattering width parameter is fixed at $K = 0.630 \, R_{\odot}$ for all cases. (1) \textbf{Dust Load}: Total number of dust particles detected in the monitoring image. (2) \textbf{Fitted Coeff.}: Global scaling factor $C$ derived from least-squares fitting. (3) \textbf{Background Noise}: RMS intensity evaluated within the polar coronal holes ($1.1$--$1.8 \, R_{\odot}$). The data are presented in the format ``$X \rightarrow Y$ ($Z\%$)'', where $X$ is the original RMS noise before correction ($I_{L1}$), $Y$ is the residual RMS noise after correction ($I_{\text{corr}}$), and the bold value $Z\%$ indicates the percentage of noise reduction. (4) \textbf{SBR}: Averaged Signal-to-Background Ratio within the inner corona ($1.05$--$1.20 \, R_{\odot}$). Similarly, the format displays the pre-correction SBR $\rightarrow$ post-correction SBR, with the bold value in parentheses indicating the multiplicative enhancement factor.}
\end{deluxetable*}

\begin{figure}[ht!]
    \centering
    \includegraphics[width=0.75\textwidth]{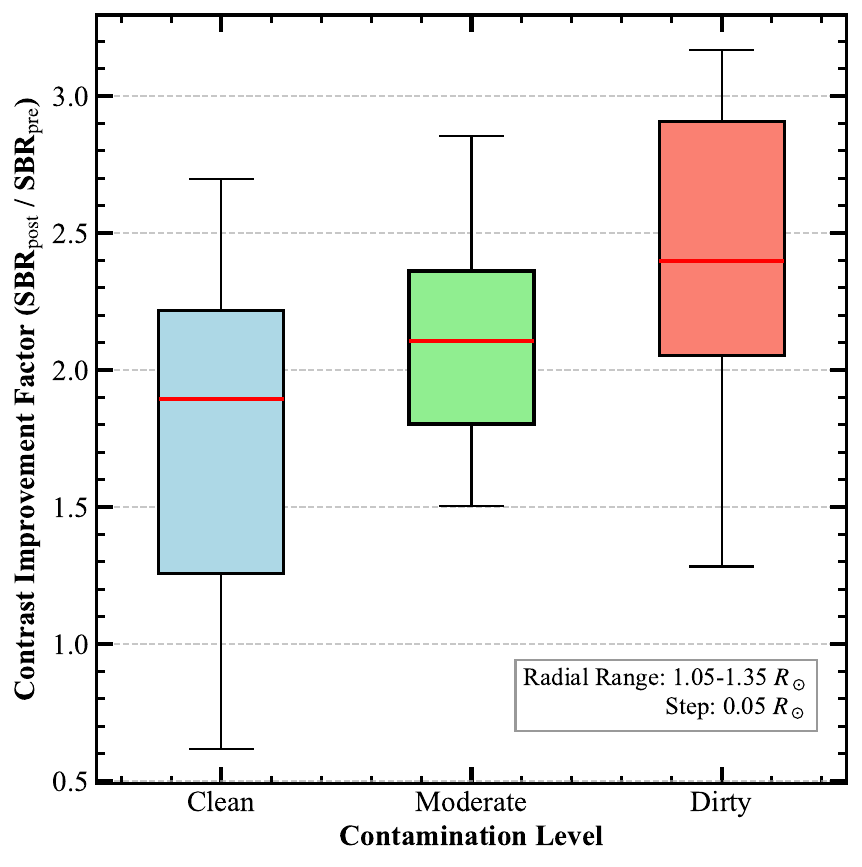}
    
    \caption{Statistical distribution of the Contrast Improvement Factor  over the extended radial range of \textbf{$1.05$--$1.35 \, R_\odot$}. 
    The CIF is calculated as the ratio of $SBR_{post}$ to $SBR_{pre}$ at discrete radial intervals with a step size of $0.05 \, R_\odot$. 
    The boxplots illustrate the performance across Clean, Moderate, and Dirty contamination levels: the \textbf{red lines} indicate the median values, the \textbf{colored boxes} represent the interquartile range , and the whiskers extend to the data range. 
    Note that despite the inclusion of the photon-noise-dominated outer region ($> 1.20 \, R_\odot$), which contributes to the spread in the distribution, the improvement factors consistently exceed 1.0 for all cases, with the ``Dirty'' case maintaining a robust median improvement of $\sim 2.4\times$.}
    
    \label{fig:Fig7}
\end{figure}

To further assess the robustness of the algorithm over an extended field of view (FOV), we expanded the analysis range to $1.05$--$1.35 \, R_\odot$. To ensure a statistically significant sample distribution, we performed radial slice sampling with a step size of $0.05 \, R_\odot$, calculating the SBR improvement factor at each altitude. Figure \ref{fig:Fig7} illustrates the statistical distribution of these sampled points across different contamination levels. The results indicate that the Contrast Improvement Factor (CIF) consistently remains above 1.0 for all cases, even with the inclusion of the outer regions ($> 1.20 \, R_\odot$) where the signal-to-noise ratio is lower.

Notably, the median values in Figure \ref{fig:Fig7} are slightly lower than the averages reported in Table \ref{tab:correction_stats}, and the data distribution exhibits a certain degree of dispersion. This behavior is physically expected: as the radial distance increases, the coronal radiation intensity decays exponentially, causing the observational data to transition from a ``stray-light-dominated'' (systematic error dominated) regime to a ``photon-noise-dominated'' regime. While the algorithm effectively removes the additive stray light pedestal, it cannot eliminate the random photon noise inherent to the sky background, which naturally imposes an upper limit on the SBR improvement in the outer FOV. Nevertheless, in the ``High Contamination'' (Dirty) case, the median improvement factor across the full FOV remains robust at approximately 2.4$\times$, demonstrating the effectiveness of the proposed method throughout the observable range.

Furthermore, by comparing the statistical results across different radial ranges ($1.20 \, R_\odot$ vs. $1.35 \, R_\odot$), a consistent pattern emerges: the SBR improvement factor exhibits a monotonically increasing trend with the severity of contamination (Clean $<$ Moderate $<$ Dirty). This indicates that as the dust load increases, the systematic bias induced by stray light constitutes a larger fraction of the total signal, and consequently, the correction yields a more significant marginal gain.

\subsection{Cross-Validation with Space-based SDO/AIA Observations} 
\label{subsec:cross_validation}

\begin{figure*}[ht!]
    \centering
    \includegraphics[width=0.95\textwidth]{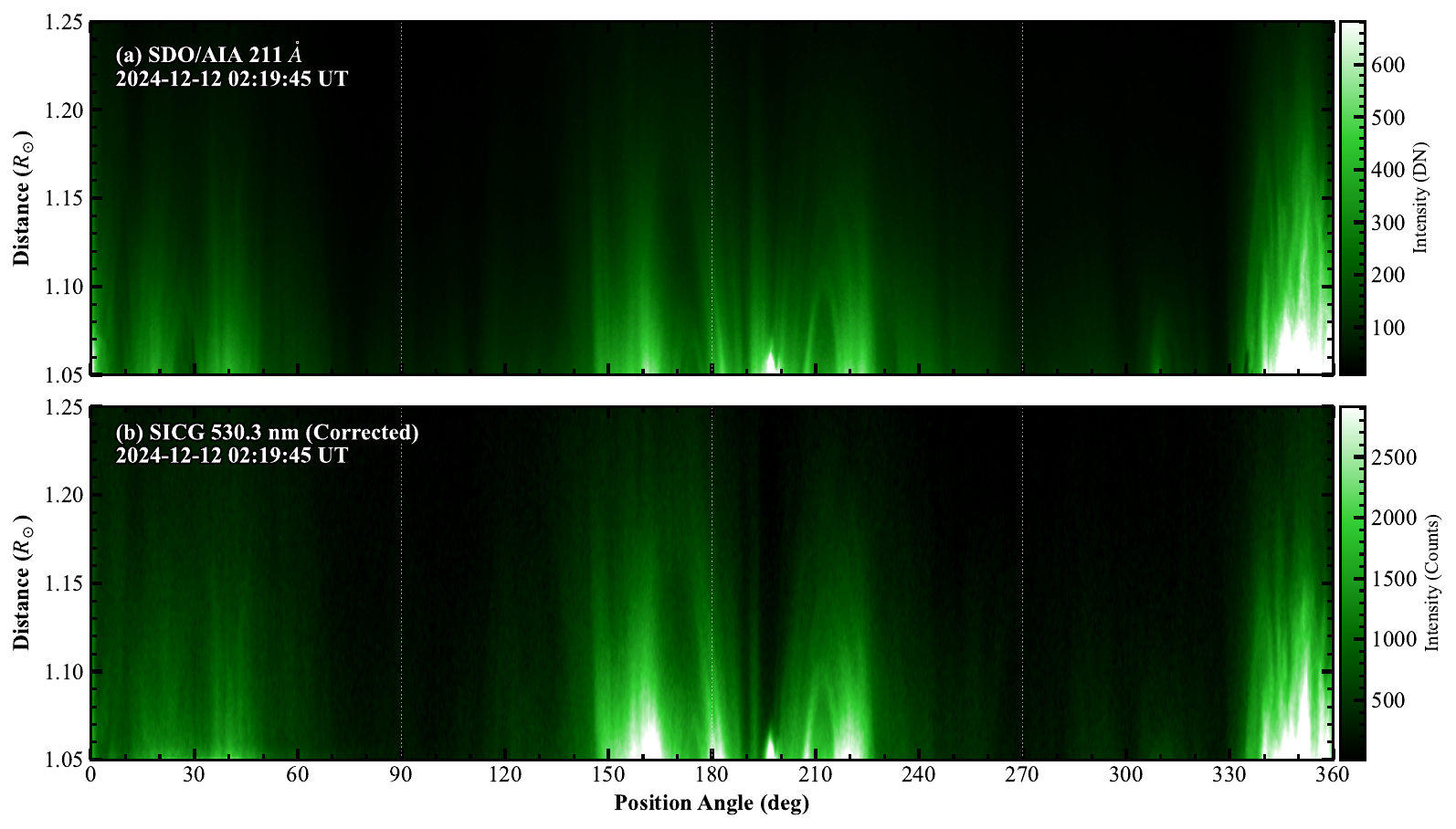}
    
    \caption{Comparison of polar-unrolled maps (Cartesian projection) between SDO/AIA 211 \AA\ and SICG 530.3 nm. 
    The maps cover the radial range of $1.05$--$1.25 \, R_\odot$ and the full azimuth ($0^\circ$--$360^\circ$). 
    \textbf{(a)} Reference image from SDO/AIA 211 \AA. 
    \textbf{(b)} Corrected SICG image for the ``High Contamination'' (Dirty) case. 
    Both images are displayed in a green colormap to highlight the structural correspondence. The vertical dashed lines mark the position angles of major streamers ($90^\circ, 180^\circ, 270^\circ$). Note the precise alignment of streamer structures and the clean background in the SICG image, validating the accuracy of the stray light removal.}
    
    \label{fig:Fig8}
\end{figure*}

Before comparing the spatial structures, it is necessary to establish the physical basis for cross-validating the ground-based visible 530.3~nm observations with the space-based EUV 211~\AA\ data. The morphological similarity between these two passbands is rooted in their shared atomic properties and emission mechanisms. First, both the 530.3~nm coronal green line and the dominant emission within the AIA 211~\AA\ passband originate from the exact same Fe~XIV ion, sharing a peak formation temperature of approximately 2.0~MK ($\log T \approx 6.3$) \citep{ODwyer2010, Lemen2012}. Consequently, they trace identical hot plasma structures, such as helmet streamers. Second, regarding the emission mechanisms, the AIA 211~\AA\ line is purely collisionally excited, with its intensity scaling as the square of the electron density ($n_e^2$). Although the 530.3~nm emission is a mixture of collisional excitation and radiative resonant scattering (which scales linearly with $n_e$), the collisional component strongly dominates in the dense inner corona (1.05--1.25~$R_\odot$) analyzed in this study \citep{Habbal2007}. Because the emission in both passbands is heavily $n_e^2$-dependent in this region, their angular intensity distributions are expected to be highly consistent. Based on this physical premise, we rigorously validated the scientific fidelity of the corrected data.

To rigorously validate the scientific fidelity of the corrected data, we performed a comparative analysis using polar-unrolled maps of the corrected SICG 530.3 nm images and simultaneous SDO/AIA 211 \AA\ observations. As illustrated in Figure \ref{fig:Fig8}, despite inherent differences in observing wavelengths and imaging mechanisms, the two datasets exhibit remarkable consistency in angular structural distribution within the range of $1.05$--$1.25 \, R_\odot$. The enhanced regions corresponding to the main equatorial streamer belts align well with the AIA results in terms of position angle, while the polar regions appear as distinct zones of low emission, consistent with typical coronal hole characteristics.

The corrected SICG images clearly resolve radial streamer structures, with angular extensions consistent with the AIA data. Furthermore, the uniform background elevation induced by stray light is effectively eliminated, resulting in a significant enhancement of contrast in dark regions. Minor discrepancies in angular brightness can be attributed to the differing sensitivities of the two passbands to plasma temperature and density. Overall, this comparison validates the effectiveness of the proposed scattering correction method in recovering the intrinsic angular structure of the inner corona, providing independent verification from space-based instrumentation.

\begin{figure*}[ht!]
    \centering
    \includegraphics[width=\textwidth]{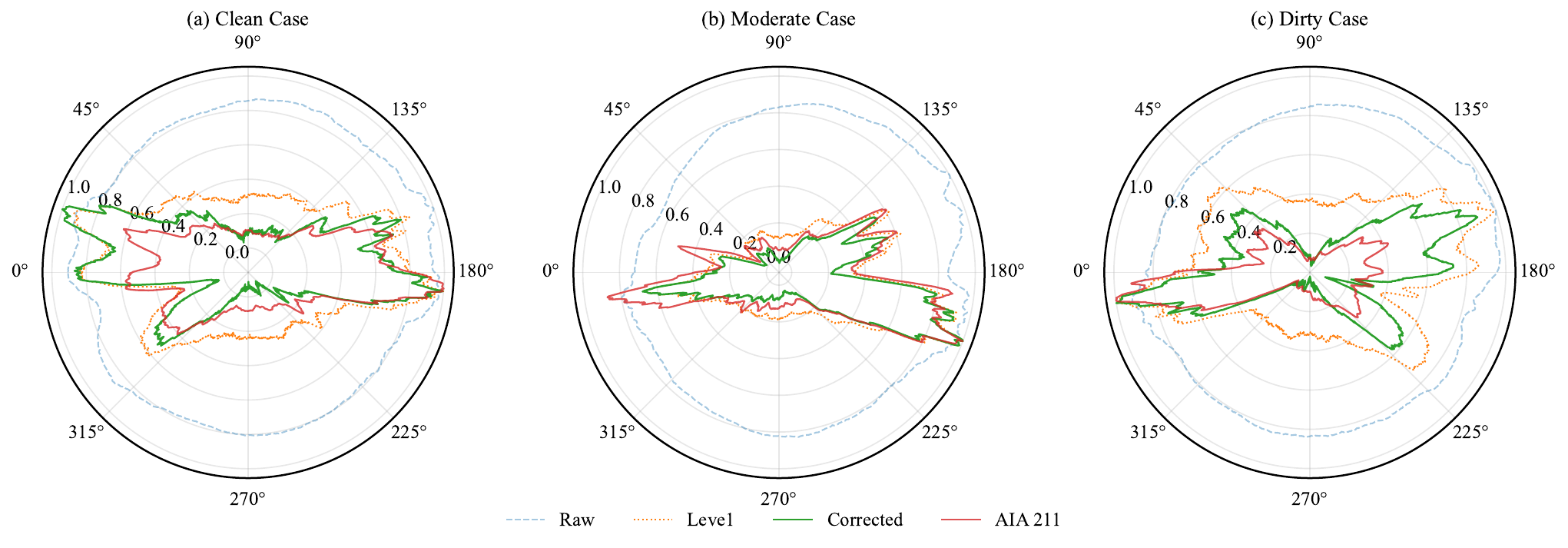}
    
    \caption{Comparison of normalized angular intensity profiles averaged over the radial range of $1.05$--$1.25 \, R_\odot$ under different contamination levels. 
    Panels \textbf{(a)--(c)} correspond to the Clean, Moderate, and Dirty cases, respectively. 
    The \textbf{gray dashed line} represents the raw SICG data ($I_{raw}$), which exhibits a flattened profile due to the strong stray light background. 
    The \textbf{orange dotted line} denotes the pre-processed \textbf{Level 1 data} ($I_{L1}$) after flat-field correction and geometric registration. 
    The \textbf{red solid line} shows the reference profile from SDO/AIA 211 \AA, indicating the true locations of coronal structures. 
    The \textbf{green solid line} presents the scattering-corrected data ($I_{corr}$). 
    All profiles are min-max normalized to facilitate direct morphological comparison. 
    The close overlap between the corrected SICG profiles (green) and the AIA profiles (red) in terms of streamer peak positions and widths strongly validates the scientific fidelity of the correction results.}
    
    \label{fig:Fig9}
\end{figure*}

To further quantitatively evaluate the structural consistency presented in the aforementioned polar-unrolled maps, we averaged the intensity radially within the selected range ($1.05$--$1.25 \, R_\odot$) to derive the angular intensity profiles and compared their characteristics across different processing stages. Figure \ref{fig:Fig9} displays these angular intensity distributions, with each curve normalized to its maximum value.

It is evident that the Raw data exhibits weak angular modulation across all three contamination levels, reflecting the quasi-isotropic background introduced by stray light. In contrast, the Corrected results show high consistency with the AIA 211 \AA\  data at the primary peak positions, successfully recovering the angular structure dominated by equatorial streamers.

As the severity of stray light contamination increases (from Clean to Dirty case), the angular contrast in the original data decreases significantly, and the intensity distribution tends to flatten. This trend is particularly pronounced in the Raw and Level 1 data, indicating that traditional pre-processing is ineffective at suppressing the impact of stray light on the angular structure. Conversely, the corrected curves maintain distinct angular modulation amplitudes under all contamination conditions, demonstrating the robustness of the proposed method against high stray light backgrounds.

Notably, despite the significant differences in stray light levels among the three cases, the morphology of the corrected angular distributions remains highly consistent in global structure and aligns well with the AIA observations. This indicates that the scattering correction method possesses excellent adaptability and stability regarding the degree of stray light contamination.

\section{Summary and Discussion}
\label{sec:discussion}

In this study, to address the critical challenge of strong additive stray light caused by objective lens dust in ground-based internally occulted coronagraphs, we proposed a correction method combining \textit{dual-path real-time monitoring} and \textit{forward physical modeling}. This method has been systematically applied to the routine data processing of the SICG. By integrating the measured dust source distribution with the instrument's optical defocus parameters, we successfully constructed and subtracted the non-uniform additive scattering background. Quantitative evaluation demonstrates that the proposed method exhibits exceptional robustness under varying contamination conditions: the polar background noise is reduced by approximately 70\% on average, and the Signal-to-Background Ratio (SBR) is improved by a factor of 2.4 to 3.7. Crucially, cross-validation with SDO/AIA observations confirms that the corrected data reproduces the morphological structures of the inner corona with high fidelity.

To place this correction method within the broader context of the overall coronal data processing pipeline, it is important to evaluate its impact on the data's overall uncertainty budget. Before correction, especially under severe contamination, the objective lens dust constitutes the overwhelmingly dominant source of systematic error. As quantitatively shown, it accounts for approximately $75\%$ of the absolute background intensity in the outer FOV and introduces extreme spatial variance (e.g., initial polar RMS of 297 counts). At this stage, the overall uncertainty budget is entirely governed by this \textit{systematic morphological distortion}, which severely degrades the Signal-to-Background Ratio (SBR $\approx 1.1$) and renders standard sky-background subtraction or absolute radiometric calibration impossible.
However, when our forward-modeling calibration is strategically applied after basic flat-fielding and before the final radiometric scaling, the polar RMS noise is consistently reduced by roughly $70\%$. This indicates that the vast majority of the dust-induced systematic bias is effectively eliminated. Consequently, the nature of the overall uncertainty budget undergoes a fundamental shift. The residual uncertainty in the corrected data ($I_{corr}$) is no longer dominated by instrumental stray light, but is instead restored to the physical \textit{statistical shot-noise} limit, dictated primarily by the irreducible Poisson fluctuations of the intrinsic atmospheric sky brightness (the remaining $\sim 71.8$~counts) and fundamental detector readout noise. By successfully removing the primary systematic barrier, this dust calibration provides the strictly required, structurally clean input for subsequent high-precision sky background subtraction and thermodynamic diagnostics in future scientific applications.

The most significant physical value of this study lies in the fact that the stray-light-corrected data successfully recovers the characteristics of \textit{hydrostatic equilibrium} in the radial coronal radiation. In the raw observations, the flattened background caused by dust scattering (decay coefficient $b \approx -9.8$) masks the true density gradient. Direct inversion using such data would lead to an artificially inflated density scale height, thereby deriving a non-physical high-temperature structure ($> 2.5$ MK). Following correction, the radial photometric profile recovers a steep decay trend ($b \approx -11.3$), which corresponds to a plasma temperature of approximately 2.0 MK. This value is in excellent agreement with the characteristic formation temperature of the SICG observing passband (\ion{Fe}{14} 530.3 nm). As noted by \citet{Wang1997}, this line primarily traces hot plasma ($\sim$2 MK) confined within closed magnetic field structures (e.g., helmet streamers), which further validates the thermodynamic fidelity of the corrected data.These results indicate that the proposed correction method is a prerequisite for precision plasma diagnostics. Only by thoroughly eliminating the systematic errors introduced by the additive background can we accurately invert the spatial distributions of electron density ($n_e$) and temperature ($T$) based on emission line intensities. This is equally critical for non-thermal velocity analysis derived from spectral line widths. As demonstrated by \citet{Huangfu2025} in their study of coronal cavities using SICG data, accurate background subtraction prevents the ``dilution'' of spectral profiles by stray light, thereby avoiding the underestimation of non-thermal line widths and associated turbulence velocities. Furthermore, given the strong physical correlation between coronal green line brightness and magnetic field intensity \citep{Zhang2022_Magnetic}, and the central role of precise spectral observations in magnetic field measurements (such as the strong/weak field techniques discussed by \citealt{Chen2023}), eliminating the non-uniform brightness bias caused by dust is of fundamental importance for constructing reliable coronal magnetic topology models. Consequently, the stray-light-corrected SICG data will provide reliable observational constraints and boundary conditions for developing more accurate coronal heating models and investigating solar wind acceleration mechanisms.

Furthermore, the significant enhancement in contrast for inner coronal observations directly benefits current frontier research regarding faint eruptive activities. The region covered by the SICG ($1.05$--$2.0 \, R_\odot$) is the critical source region for CME acceleration and flux rope formation. As demonstrated by \citet{Tian2013} using CoMP observations, ground-based spectral coronagraphs possess a unique field-of-view advantage in capturing the early kinematics of CMEs; however, this region is also typically the most severely affected by stray light contamination. Our results show that the SBR is improved by a factor of more than 3 after correction, implying that nascent CME bubbles, cavities, and fine-scale mini-streamers, which were previously submerged in the ``stray light haze,'' are rendered detectable. In particular, for the analysis of fine-scale structures during the early evolution of CMEs, such as the ``bright front'' and its physical nature recently investigated by \citet{song2025}, high-contrast inner coronal observational data are critical for distinguishing the shock front from the driver ejecta. Eliminating stray light interference will directly improve the identification accuracy of such diffuse features. Regarding coronal seismology, accurate background subtraction is a prerequisite for extracting weak Alfv\'{e}n wave or fast magnetosonic wave perturbation signals. Eliminating the non-uniform dust background can significantly reduce spurious signals in time-domain analysis, thereby improving the signal-to-noise ratio and accuracy of wave mode identification.

From the perspective of observational methodology, this study demonstrates a \textit{paradigm shift} in treating scattering from objective lens dust: moving from regarding it as ``uncontrollable noise'' to handling it as a ``modelable physical component.'' By introducing deterministic monitoring information and rigorous forward optical modeling, the dependence of ground-based coronal observations on stringent physical cleanliness is significantly reduced, thereby increasing the effective observational duty cycle of the instrument. This strategy provides a generalizable technical pathway for the stray light suppression design of next-generation large-aperture ground-based coronagraphs.

It is important to note that the proposed algorithm focuses on removing instrumental stray light caused by objective lens dust. The residual background in the corrected images (represented by the constant term $c$ in the exponential fit $I = A \cdot \exp[b(r-1)] + c$) is primarily attributable to atmospheric scattering (sky brightness) and F-corona radiation. Further separation of these components requires the incorporation of polarimetric observations or long-timescale statistical methods. Additionally, in extremely rare scenarios of high-density dust coverage, the assumption of linear incoherent superposition may introduce secondary errors, which will be further evaluated in future work.

In summary, the stray light correction method proposed in this paper has successfully elevated the quality of SICG ground-based observational data to a level suitable for precise physical studies of the inner corona. It enables the SICG to provide a critical complement to space-based observations within this key height range and offers reliable observational constraints for investigating the physical mechanisms underlying coronal heating and solar wind acceleration.

\begin{acknowledgments}
 We sincerely thank the anonymous referee for the careful reading and highly constructive comments, which have significantly improved the clarity and physical rigor of this manuscript. We acknowledge the Chinese Meridian Project for providing high-quality data from the SICG. We also thank the NASA/SDO and the AIA science team for open data access. We sincerely thank the NAOJ team for the long-term 10~cm coronagraph collaborations, and acknowledge the data resources from the National Space Science Data Center, National Science and Technology Infrastructure of China (\url{http://www.nssdc.ac.cn}). 

This work is supported by the National Natural Science Foundation of China (NSFC grant Nos.~12173086, 12373063, 11533009, 12163004, 12473089, and 42274227), the Yunnan Fundamental Research Projects (grant Nos.~202501AS070004 and 202401AT070140), and the Yunnan Key Laboratory of Solar Physics and Space Science (grant No.~202205AG070009).
\end{acknowledgments}

\bibliographystyle{aasjournal} %
\newpage 
\bibliography{myrefs}

\end{document}